\documentclass[preprint]{aastex}

\shorttitle{Nearby stars from the NLTT: SAAO}

\shortauthors{Reid {\it et al.}}

\begin {document}
\title{Meeting the Cool Neighbours, II: Photometry of southern NLTT stars}

\author {I. Neill Reid}
\affil {Space Telescope Science Institute, 3700 San Martin Drive, Baltimore,
MD 21218; \\
Department of Physics and Astronomy, University of Pennsylvania, 209 South 33rd Street,
Philadelphia, PA 19104
e-mail: inr@stsci.edu}

\author {D. Kilkenny}
\affil {South African Astronomical Observatory, PO Box 9, Observatory 7935, South Africa}

\author {K. L. Cruz}
\affil {Department of Physics and Astronomy, University of Pennsylvania,  209 South 33rd Street,
Philadelphia, PA 19104}

\begin{abstract}

We present BVRI photometry of 180 bright, southern nearby-star candidates.
The stars were selected from the New Luyten Two-Tenths proper motion catalogue based
on optical/infrared colours, constructed by combining Luytens's m$_r$ estimates with
near-infrared photometry from the 2-Micron All Sky Survey. Photometric parallaxes
derived from (V-K$_S$), (V-I) and (I-J) colours, combined with the limited available
astrometry, show that as many as 108 stars
may lie within 20 parsecs of the Sun. Of these, 53 are new to nearby star
catalogues, including three within 10 parsecs of the Sun. 

\end{abstract}


\section {Introduction}

This is the second in a series of papers which present the results of our
search for previously unrecognised stars within the immediate Solar Neighbourhood. As discussed in
Paper I (Reid \& Cruz, 2002), the availability of large-scale sky surveys at near-infrared
wavelengths, notably the 2-Microns All Sky Survey (2MASS: Skrutskie {\sl et al.}, 
1997), in combination with published
catalogues and Schmidt-based photographic observationss at optical wavelengths, 
has greatly enhanced 
our capabilities for detecting the low-luminosity main-sequence stars and brown dwarfs. 
The primary goal of our project, undertaken under the auspices of the NASA/NSF NStars initiative, 
is the identification of all late type dwarfs (concentrating on spectral types M and L) within 
20 parsecs of the Sun. 

As a first step towards identifying the Sun's nearest neighbours, we have used 2MASS data
to enhance an old technique, cross-referencing the infrared catalogue against proper motion
stars in the New Luyten Two-Tenths Catalogue (Luyteon, 1980 - NLTT). Once correlated, we can use the 
($m_r-K_S$) colour as a crude photometric parallax estimator, with the long baseline compensating
to some extent for the uncertainties in the red magnitudes. 
Paper I describes the definition of an initial sample of nearby-star candidates, drawn from
NLTT sources which have potential 2MASS counterparts within a 10-arcsecond search radius.
By applying a series of cuts in colour-magnitude and colour-colour planes, we reduced the list
of 23795 optical/IR matches to 1245 sources with photometric properties consistent with their being
late-type dwarfs within 20 parsecs of the Sun. These stars constitute NLTT Sample 1. Paper I
compiles photometric data from the literature for 469 stars, and uses colour-magnitude relations
at (M$_V$, (V-K), (M$_V$, (V-I)) and (M$_I$, (I-J)) to estimate distances to those stars. Three
 hundred of those stars, and a further 39 ultracool (spectral type $>$M6) dwarfs, have formal 
distances of less than 20 parsecs, including 76 stars not previously included in nearby star catalogues.

The current paper continues analysis of the NLTT Sample 1, presenting optical photometry of a
sample of 180 relatively-bright southern stars. The following section outlines the sample and presents
the observations. Section 3 describes our procedures for estimating distances to these stars, and Section
4 discusses some of the more interesting stars in the sample. Our results are
summarised in the final section.

\section {Observations}

\subsection {The sample}

As described in Paper I, the 1245 stars in NLTT Sample 1 were selected on the basis of their
having locations in the ($m_r$, ($m_r$-K$_S$)) and ((J-H)/(H-K)) planes consistent 
with mid- or late-type M dwarfs within 20 parsecs of the Sun. Regions within $\pm10^o$ of
the Galactic Plane were excluded {\sl a priori}, since the NLTT catalogue has a
significantly brighter limiting magnitude at those latitudes. The selected stars have
magnitudes in the range $8 < m_r < 20$, with over 70\% lying between 11th and 16th
magnitude. They span the full range of Right Ascension, 
although the majority lie at northern Declination, reflecting both
the areal coverage of the second incremental release of 2MASS data and incompleteness 
in the NLTT south of $\delta = -30^o$. Several hundred stars, 
however, lie south of the equator. 

Southern hemisphere proper-motion stars have generally received less attention than
their northern counterparts, and, as a result, even relatively bright objects in
the current sample have no previous detailed measurements. Figure 1 shows
the distribution on the celestial sphere and in the ($m_r$, ($m_r$-K$_S$)) colour-magnitude
plane of the 180 NLTT dwarfs targeted here. 

\subsection {Photometry}

All of the data were obtained (by DK) between February and July 2001, 
using the St. Andrews photometer with a Hamamatsu R943-02 GaAs photomultiplier
on the 1.0-metre telescope at the Sutherland station of the South African Astronomical
Observatory. Most observations were made through a 21-arcsecond diameter, though a 
31-arcsecond diameter aperture was employed for poorer seeing. As discussed further
below, the relatively large apertures sometimes lead to our photometry including
a contribution from other stars at small angular separation.

The observations were made using a Johnson-Cousins BVRI filter set and reduced
using standard techniques (Kilkenny {\sl et al.}, 1998). 
Only VRI measurements were made for most of the fainter stars. In addition to 
E-region standard stars (Cousins, 1973; Menzies {\sl et al.}, 1989), we also made
substantial numbers of calibrating observations of red standards from the list of
Kilkenny {\sl et al.} (1998), which extends the Cousins BVRI system into the
(approximate) range $2 <$ (V-I) $< 3$. Beyond (V-I)=3, we (and
others) are extrapolating colour equations, but observations of Proxima Centauri compared
to Bessell (1990) indicate we are close to his system. In any case, few stars amongst the
present targets  have (V-I)$> 3.$

The results from our photometry are listed in Table 1, where we also give
the positions and near-infrared data from the 2MASS catalogue.
We were able to obtain two or more observations of approximately half the sample. Those
repeated measurements indicate that the standard deviations for a single measurement are
typically $\sigma_V = 0.015$ mag., $\sigma_{B-V} = 0.011$,  $\sigma_{V-R} = 0.008$
and  $\sigma_{V-I} = 0.013$. Figure 2 compares the VRI and JHK two-colour diagrams for
our dataset against observations of known nearby stars from Bessell (1990) and Leggett (1992), where 
we have supplemented the optical observations from the former reference with 2MASS data. We also
show the dwarf and giant (H-K)/(J-H) sequences from Bessell \& Brett (1988), transforming the
latter to the 2MASS system following Carpenter (2001). As discussed in Paper I, Leggett's CIT
near-infrared data closely match the 2MASS system. There is excellent agreement
between our observations and the standard sequences, showing that the optical and near-infrared
photometry are internally consistent.

Figure 3 compares  optical/near-infrared colour-colour diagrams for our sample against
data for nearby stars. The majority of the NLTT stars are in excellent agreement with the
standard sequences, but there are a few outliers in both (B-V) and (V-I). As noted above,
our optical data are derived from aperture photometry, and in some cases, identified in Table 1,
the aperture included additional stars besides the program object. While in many cases the optical
photometry is not significantly affected, most of the outliers in the BVK and VIK planes are 
amongst those stars. B-band observations are particularly susceptible, since our program stars are
faintest at those wavelengths, and we have excluded several measurements where the (B-V) colour was
clearly incompatible with the VRI data. Finally, G 163-4 (LHS 2297) lies only 16 arcseconds from its
brighter common proper motion companion, LHS 2296. Comparison between our data and photometry
by Weis (1996) suggests that the latter star has contaminated our measurements, and we adopt
the magnitudes listed by Weis for this star.

Four of the five outliers in the VIK plane are known binaries
where the components have similar magnitudes. All four are resolved by 2MASS, so the (V-K$_S$) colour
is therefore (V(AB)-K$_S$(A)). The fifth outlier in VIK is LP 779-34, which is also an
outlier in BVK. LP 779-33 is listed as a common proper motion companion in the NLTT, but is
several arcminutes distant (and bluer than our (m$_r$-K$_S$) limit). The 
location of LP 779-34 on the two-colour diagram probably reflects the contribution 
of a nearby, similar-magnitude field star
to the SAAO photometry. The measured VRI colours suggest that the field star is also an
M dwarf. We have taken the composite nature of the optical photometry for all of these 
stars into account in computing the photometric parallaxes given in Section 3. 

\subsection {Comparison with previous observations}

A number of stars observed in the course of our present program are well-known
nearby stars, and have published broadband photometry. In particular, 26
stars were observed by Bessell (1990) in his survey of late-type dwarfs in the
second Catalogue of Nearby Stars (Gliese, 1969; Gliese \& Jahrei{\ss}, 1979), while
10 are included in Leggett's (1992) compilation of optical and near-infrared photometry
and three were observed by the RECONS group (Patterson {\sl et al.}, 1998). 
A further 31 are amongst the proper motion stars observed by Weis (1991, 1993, 1996), and,
finally, 15 stars have photometry by Eggen (1987). The latter two sets of observations are
on the Kron RI system, but we have used the transformations given by Bessell \& Weis (1987)
to transform to Kron-Cousins magnitudes. As noted above, we adopt Weis' data for G 163-4.

Table 2 and Figure 4 show the statistical results from a comparison of our observations and data
from the literature. We have excluded one observation from this comparison: Bessell lists V=10.51 for
Gl 386, while Weis lists V=10.97. Our photometry agrees with the latter measurement, so we 
omit the former. 
The main discrepancies lie with Eggen's photometry, partly because his V-band measurements 
are given to the nearest 0.1 magnitude, and may in some cases be estimates based on the
(R-I) colours.  Overall, the agreement is consistent with our
internal estimate of the photometric uncertainties, and there
is no evidence for any significant residual colour terms. 

\section {Distance estimates}

Our goal in this project is the identification of stars likely to lie within 20 parsecs
of the Sun. The NLTT  sample discussed in this paper
includes seventy-one stars already identified as such in
the preliminary version of the third Catalogue of nearby Stars (Jahrei{\ss} \& Gliese, 1991: pCNS3). 
However, while all 180 stars  are relatively bright, only 56 stars have trigonometric parallax 
measurements. As a result, we must rely on photometric parallax as our primary method
of estimating distances. We have followed an identical approach to that outlined in paper I, 
computing photometric parallaxes from the observed (V-I), (I-J) and (V-K$_S$) colour indices
using the polynomial calibrations outlined in that paper; combining those estimates
to give an averaged photometric distance estimate; and deriving our final distance
estimate through a weighted average with the trigonometric measurement, should such exist.
Following the discussion outlined in Paper 1, we set a lower limit of $\pm0.3$ mag. to the 
weight assigned to the photometric distance modulus, and we note
that the trigonometric parallax offers the best estimate of the distance to an individual star.
The results are listed in Table 3, where we give the uncertainty associated with the
individual measurements. As discussed above and in the notes to the Table, we make due
allowance for known binaries with joint optical photometry. Table 3 includes several
wide common proper motion pairs, notably LP 890-44/45, -25:10333A/B and CD -44:836/LP 993-116.
In each case, the photometric parallaxes of both components agree within the formal
uncertainties.

Figure 5 plots, as a function of (V-I) colour,  the residuals between the individual 
distance modulus estimates and the final averaged value. 
The larger residuals, and systematic
trend in $\delta$(av-(I-J)),  near (V-I)=2.9 reflect the sharp steepening of the
main-sequence, and consequent larger uncertainties, at that colour (see Paper I). 
Comparing the trigonometrically-based distance
modulus estimates against the averaged photometric parallax results give
\begin{displaymath}
\langle (m-M)_{\pi} \ - \ (m-M)_{ph} \rangle \ = \ 0.01 \pm 0.81 mag.
\end{displaymath}
Restricting the comparison to the 42 stars with $\pi_{trig}$ measured to an accuracy of
10\% or better gives
\begin{displaymath}
\langle (m-M)_\pi \ - \ (m-M)_{ph} \rangle \ = \ 0.06 \pm 0.48 mag.
\end{displaymath}
As Figure 5 shows, there is no evidence for significant systematic bias, and the 
residuals are broadly consistent with the expected uncertainties in the photometric 
parallax calibrations (Paper I). 
Table 3 gives our distance estimates for each star and identifies those likely to fall
within the 20-parsec distance limit. Figure 6 plots the distance distribution 
as a function of (V-I) colour. 

\section {Discussion}

Table 1 lists photometry for 180 NLTT systems, including at least four binary systems. Based on
our distance estimates, 86 of those systems, including three binaries (Ross 948, LP 675-76/77
and LP 726-11/12) have distances of less than 20 parsecs, while a further 28 lie within 1$\sigma$
of the 20-parsec boundary. Sixty-one of the 114 systems are included in the pCNS3, 
but the remaining
48 are additions to nearby star catalogues (although some are listed at the pre-CNS4
website, {\sl http://www.ari.uni-heidelberg.de/aricns/}).
The additional stars are identified in the final column 
of Table 3. Moving the distance limit to 25 parsecs, the value adopted
in the pCNS3, embraces 132 systems, 65 of which are not included in that catalogue. Given the
relatively bright magnitudes of the stars in our sample, this emphasises the incompleteness 
of nearby star surveys in the southern hemisphere. 

Considering the immediate Solar Neighbourhood, our distance estimates place 14 of the 180 stars 
within 10 parsecs of the Sun. Those systems include Gl 84, 190, 357, 382, 628, GJ 1065 and
LHS 1731, all of which have trigonometric parallaxes which exceed 0.1 arcseconds. Four other 
stars are listed in the pCNS3: Gl 540.2 and LHS 2520, 1723 and 2836. The three new
identifications are LP 993-116 (M$_V = 13.45$),  
LHS 6167 (M$_V = 14.63$) and G 161-71 (M$_V = 14.76$). All three systems have formal distance estimates
between 6 and 7 parsecs. As noted above, LP 993-116 is identified in the NLTT as a
common proper motion companion, separation 44 arcseconds, of CD -44:836, an M5 dwarf. The
latter star is in our sample, and we derive an estimated distance of 10.5 parsecs. 
Given the individual uncertainties, the agreement is reasonable. 

There are currently approximately
250 systems known with 10 parsecs of the Sun. Thus, the 5 new identifications listed in this
paper and Paper I represent an increase of only $\sim2\%$ in the inferred local stellar space density.
However, these additions are drawn from a relatively well-studied subset of our NLTT candidates.
Combined with the literature data discussed in Paper I, we have optical photometry, and photometric
parallaxes, for 649 of the 1245 stars in NLTT Sample 1, while a further 39 ultracool dwarfs have
distance estimates based on either optical spectroscopy or (J-K$_S$) colours.  
Four hundred and fifty-six of those stars are identified as potentially within 20 parsecs of the 
Sun. Subsequent papers in this series will provide distance estimates for the remaining
596 stars in NLTT Sample 1, as well as extending coverage to include NLTT dwarfs which lack
2MASS counterparts within 10 arcseconds of the nominal position.

\acknowledgements  
The authors thank the anonymous referee for a number of extremely helpful comments.
This NStars research was supported partially by a grant awarded as part of the NASA Space 
Interferometry Mission Science Program, administered by the Jet Propulsion Laboratory, Pasadena.
This publication makes use of data products from the Two Micron All Sky Survey, which is
a joint project of the University of Massachusetts and the Infrared Processing and Analysis 
Center/California Institute of Technology, funded by the National Aerospeace and Space Administration
and the National Science Foundation. We acknowledge use of the NASA/IPAC Infrared Source Archive (IRSA), 
which is operated by the Jet Propulsion Laboratory, California Institute of Technology, under contract with
the  National Aerospeace and Space Administration.
We also acknowledge making extensive use of the SIMBAD database, maintained by Strasbourg Observatory,
and of the ADS bibliographic service. 
We would like to thank the Time Assignment Committee of the South African Astronomical
Observatory for the allocation of observing time for this project.

\begin{deluxetable}{rrrrrrrrrrcrrrrrc}
\rotate
\tabletypesize{\scriptsize}
\tablecolumns{16}
\tablenum{1}
\tablewidth{0pt}
\tablecaption{Photometry and astrometry of NLTT stars}
\tablehead{
\colhead{NLTT} & \colhead {Name} & \colhead {LHS} &
\colhead{$\alpha$ (2000)} & \colhead {$\delta$} & \colhead {m$_r$} &
\colhead{V } & \colhead {B-V } & \colhead {V-R } &
\colhead{V-I } & \colhead {n$_{obs}$ } &\colhead {J } &
\colhead{H } & \colhead {K$_S$ } & \colhead {$\pi_{trig}$(mas) } & \colhead {Ref$_\pi$ } }
\startdata
 -21:368    & G272-145   &6041& 02 \ 03 \ 20.7&-21 \ 13 \ 42& 11.0& 11.16&  1.45&  1.02&  2.31&1& 7.617& 7.031& 6.772&  33.9$\pm$22.6 &  2 \\ 
 649- 30     &           &   & 02 \ 03 \ 30.2& -04 \ 54 \ 41& 11.0& 11.11&  1.46&  0.91&  1.83&2& 8.199& 7.533& 7.354&  34.9$\pm$ 2.4 &   1 \\ 
 -18:359    & Gl  84     & 149& 02 \ 05 \ 04.7&-17 \ 36 \ 52& 10.0& 10.20&  1.52&  1.05&  2.34&1& 6.537& 5.923& 5.686& 105.9$\pm$ 2.0 &   1 \\ 
 709- 40     &    &1347& 02 \ 09 \ 36.0&-14 \ 21 \ 32& 11.1& 11.74&  1.50&  1.05&  2.34& 1& 8.136& 7.498& 7.256&  50.3$\pm$ 3.3 &   1 \\ 
 709- 43     &           &   & 02 \ 10 \ 03.6& -8 \ 52 \ 59& 12.0& 13.05& \nodata&  1.14&  2.65&2& 8.951& 8.362& 8.092& 70.2$\pm$16.3 & 2 &     \\ 
 885- 38    &  &1376& 02 \ 16 \ 35.0&-30 \ 58 \ 07& 11.7& 13.10& \nodata&  1.22&  2.84&2& 8.770& 8.233& 7.854& \nodata &     \\ 
 941- 38     &           &1380& 02 \ 17 \ 56.7&-35 \ 37 \ 00  & 11.3& 11.63&  1.49&  0.96&  2.01&1& 8.507& 7.933& 7.672&  36.2$\pm$ 2.7 &   1 \\ 
 829- 51     &           &   & 02 \ 23 \ 24.0&-23 \ 18 \ 24& 13.5& 14.21& \nodata&  1.23&  2.83&2& 9.861& 9.318& 9.005& \nodata &     \\
\enddata
\tablecomments{ 
Column 1 lists the designation from the NLTT catalogue: R=Ross, W=Wolf, Oxf=Oxford catalogue. We
have added Lowell Observatory identifications (Giclas {\sl et al.}, 1971). 
Column 2 lists an alternative name, usually from the pCNS3; 
Column 3 gives the LHS number; 
Columns 4 and 5 list the position of the 2MASS source; 
Column 6 lists $m_r$ from the NLTT, and 
columns 7-10 list the optical photometry; column 11 gives the number of observations; 
Columns 12-14 list the 2MASS photometry; 
Column 15 lists the trigonometric parallax, if available, and column 16
gives the source of the astrometry: 1 - Hipparcos (ESA, 1997); 2 - Yale parallax
catalogue (van Altena {\sl et al.}, 1995). \\
Notes for individual stars : 
$^1$ - Two stars in aperture for optical photometry; in most cases, the second star is too faint
to affect the VRI data. \\
$^2$ - possible variable. \\
$^3$ - Joint optical photometry for known binary:
Ross 948/LP 735-11 - $m_r$(A)=11.0, $m_r$(B)=11.3, $\Delta = 4$ arcsec; 
LP 793-24/25 - $m_r$(A)=11.9, $m_r$(B)=12.2, $\Delta = 4.5$ arcsec; 
LP 675-76/77 - $m_r$(A)=12.7, $m_r$(B)=13.0, $\Delta = 8$ arcsec; 
LP 726-11/12 - $m_r$(A)=12.7, $m_r$(B)=13.6, $\Delta = 8$ arcsec. \\
$^4$ - Photometry from Weis (1996). 
} 
\end{deluxetable}

\newpage

\begin{table}
\begin{center}
{\bf Table 2 } \\
{Photometry: external comparison}
\begin{tabular}{rcrrrcrc}
\tableline\tableline
Reference & & V & (V-R) & (V-I) & N & (B-V) & N \\
\tableline
Weis & $\Delta$ & -0.010 & -0.006 & -0.005 & 33 & -0.010 & 23 \\
     & $\sigma$&  0.033 &  0.018 &  0.024 &    &  0.019 &  \\
Bessell & $\Delta$ & -0.012 & -0.003 & -0.001 & 25 & -0.014 & 25 \\
     & $\sigma$&  0.022 &  0.009 &  0.014 &    &  0.017 &  \\
Eggen & $\Delta$ & -0.039 & -0.037 & -0.067 & 15 & 0.006 & 1 \\
     & $\sigma$&  0.106 &  0.082 &  0.139 &    &   &  \\
Leggett & $\Delta$ & 0.002 & 0.004 & 0.013 & 10 & -0.024 & 8 \\
     & $\sigma$&  0.012 &  0.008 &  0.009 &    &  0.009 &  \\
Patterson {\sl et al.} & $\Delta$ & 0.019 & -0.016 & 0.014 & 3 \\
     & $\sigma$&  0.025 &  0.013 &  0.034 &    \\
\tableline\tableline
\end{tabular}
\end{center}
Notes: \\
Residuals are given in the sense V$_{SAAO}$ - V$_{other}$ 
\end{table}
\begin{deluxetable}{rcccccrrrr}
\rotate
\tabletypesize{\scriptsize}
\tablecolumns{10}
\tablenum{3}
\tablewidth{0pt}
\tablecaption{Distance Modulus Estimates for NLTT stars}
\tablehead{
\colhead{NLTT} & \colhead {(m-M)$_{V-K}$} & \colhead {(m-M)$_{V-I}$} &
 \colhead {(m-M)$_{I-J}$} & \colhead {(m-M)$_{ph}$} &\colhead {(m-M)$_\pi$} & 
\colhead {d$_f$(pc)} & \colhead{M$_K$} & \colhead{M$_V$} & \colhead{ 20pc?} } 
\startdata
-21:368     &   0.76$\pm$0.41&   0.63$\pm$0.40&   0.94$\pm$0.42&   0.77$\pm$0.24&   2.35$\pm$2.39&   15.5$\pm$2.0&   5.82&  10.21& Y   \\ 
 649- 30     &   2.15$\pm$0.41&   2.21$\pm$0.40&   2.13$\pm$0.42&   2.16$\pm$0.24&   2.29$\pm$0.15&   28.1$\pm$1.7&   5.11&   8.87& N   \\ 
 -18:359     &  -0.51$\pm$0.41&  -0.45$\pm$0.40&  -0.50$\pm$0.42&  -0.49$\pm$0.24&  -0.12$\pm$0.04&    9.3$\pm$0.2&   5.85&  10.37& Y   \\ 
 709- 40     &   1.11$\pm$0.41&   1.10$\pm$0.40&   1.34$\pm$0.42&   1.18$\pm$0.24&   1.49$\pm$0.15&   19.0$\pm$1.1&   5.87&  10.35& Y   \\ 
 709- 43     &   1.20$\pm$0.41&   1.41$\pm$0.40&   1.34$\pm$0.42&   1.32$\pm$0.24&   0.77$\pm$0.57&   16.8$\pm$1.9&   6.96&  11.92& Y+  \\ 
 885- 38     &   0.50$\pm$0.41&   0.47$\pm$0.75&   0.07$\pm$0.70&   0.37$\pm$0.32&   \nodata       &   11.9$\pm$1.6&   7.48&  12.73& Y   \\ 
 941- 38     &   2.24$\pm$0.41&   2.12$\pm$0.40&   2.30$\pm$0.42&   2.22$\pm$0.24&   2.21$\pm$0.17&   27.7$\pm$1.8&   5.46&   9.42& N   \\ 
 829- 51     &   1.72$\pm$0.41&   1.62$\pm$0.75&   1.18$\pm$0.70&   1.54$\pm$0.32&   \nodata       &   20.4$\pm$2.8&   7.46&  12.66& ?+  \\
\enddata
\tablecomments{ 
Column 1 lists the designation from the NLTT catalogue, adding Lowell Observatory identifications; \\
Column 2 gives the distance modulus derived from the (V-K$_S$) photometric parallax; Columns 3 and 4 list
distance moduli based on (V-I) and (I-J), respectively, for stars with I-band photometry; Column 5
gives the weighted average of the photometric parallax measurements; Column 6 lists the
distance modulus indicated by the trigonometric parallax; \\
Column 8 gives our final estimate of the distance, based on a weighted average of the photometric
average and the trigonometric result; \\ Columns 9 and 10 list the resultant absolute magnitudes
at K and V, respectively; \\ Column 11 indicates whether the star lies within our distance limit of
20 parsecs (Y), within 1$\sigma$ of the boundary (?) or beyond the limit (N).A (+) indicates that the
star is not included in the pCNS3. \\
$^1$ - binary stars where we have joint photometry for both components. We have used the 
NLTT magitudes to estimate the relative contribution, and the distance and absolute magnitrudes are
based on the estimated optical/near-infrared colours of the primary. }
\end{deluxetable}

\newpage

\begin{figure}
\plotone{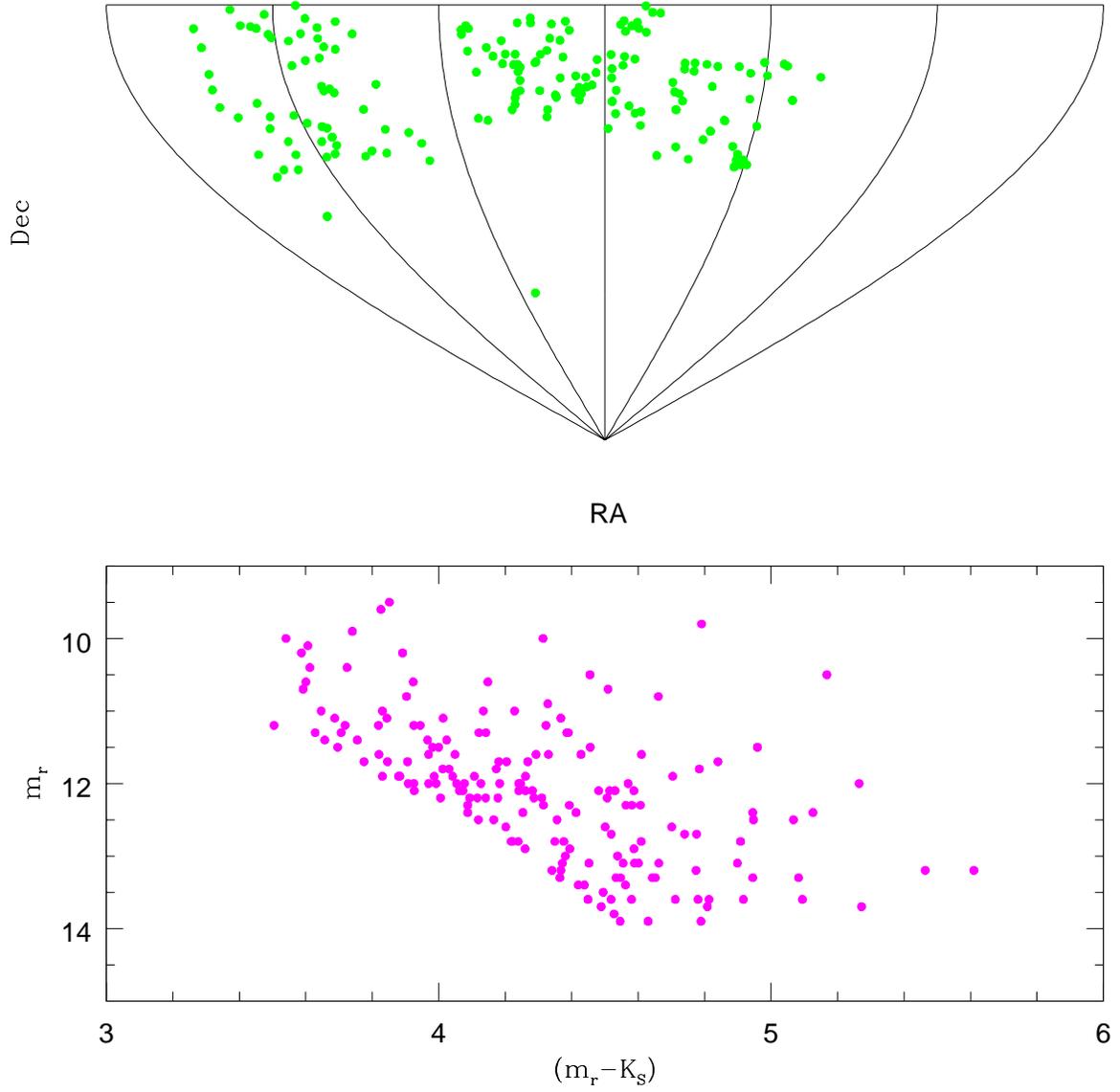}
\caption{ The ($\alpha$, $\delta$) and colour-magnitude distributions of the stars
in the present sample. The grid lines in the Aitoff projection in the upper diagram
are plotted at a spacing of 4 hours in Right Ascension. The lower diagram combines 
photometry from the NLTT catalogue and the 2MASS database. } 
\end{figure}

\begin{figure}
\plotone{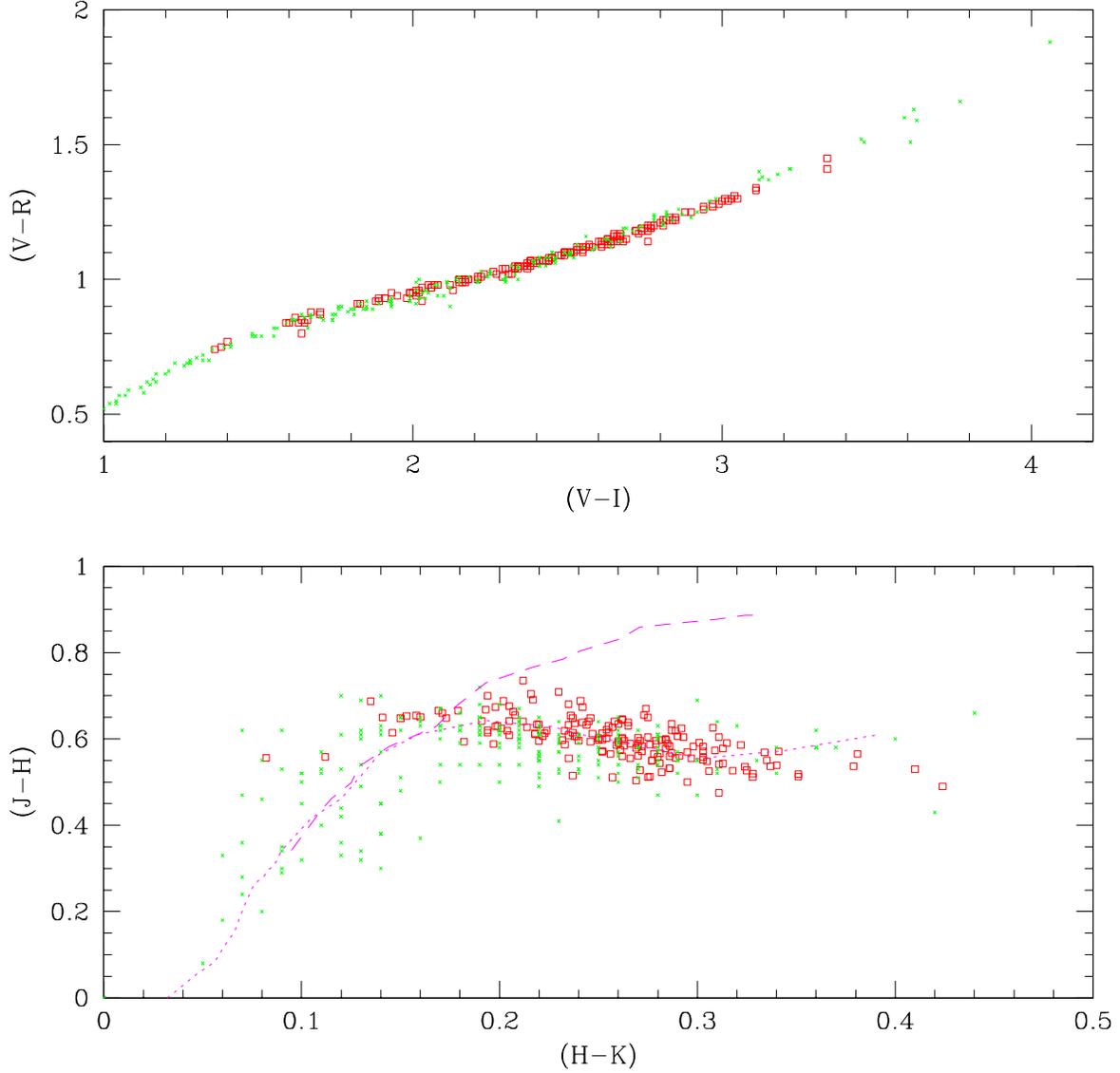}
\caption{ (V-R), (V-I)) and ((J-H), (H-K)) two-colour diagrams. NLTT stars from Table 1 are
plotted as open squares, while crosses plot data for nearby stars with accurate trigonometric
parallaxes. Photometry for the latter stars is from Bessell (1990) and Leggett (1992), supplemented,
where necessary, by 2MASS near-infrared data. The dotted line in the JHK diagram marks the mean relation
for main-sequence stars, while the dashed line plots the giant sequence (from Bessell \& Brett, 
1988, transformed to the 2MASS system following Carpenter, 2001). }
\end{figure}

\begin{figure}
\plotone{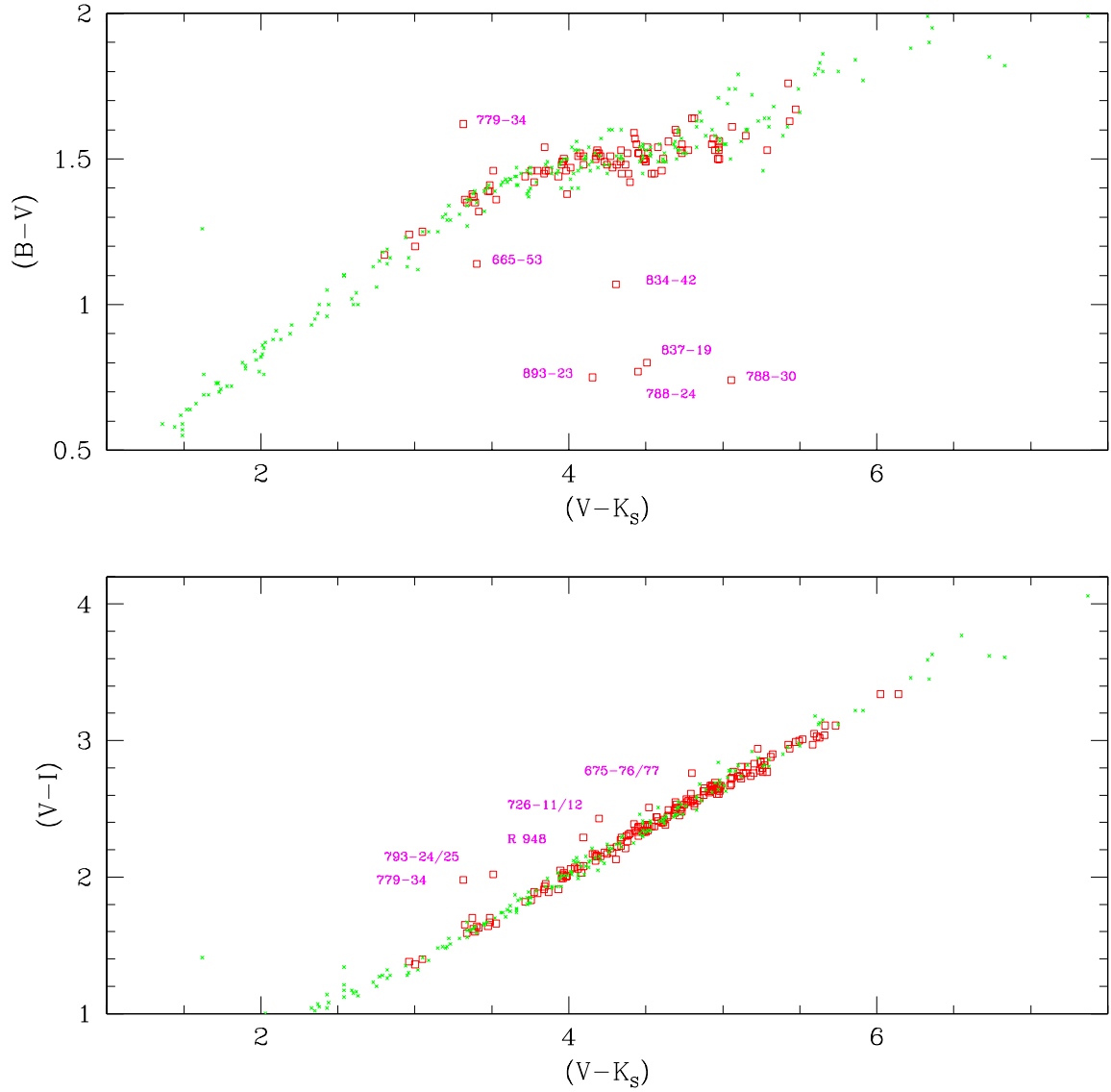}
\caption{ Optical/near-infrared colour-magnitude diagrams for the stars listed in Table 1. 
As in Figure 2, crosses are data for known nearby stars. The outliers are discussed in the text.}
\end{figure}

\begin{figure}
\plotone{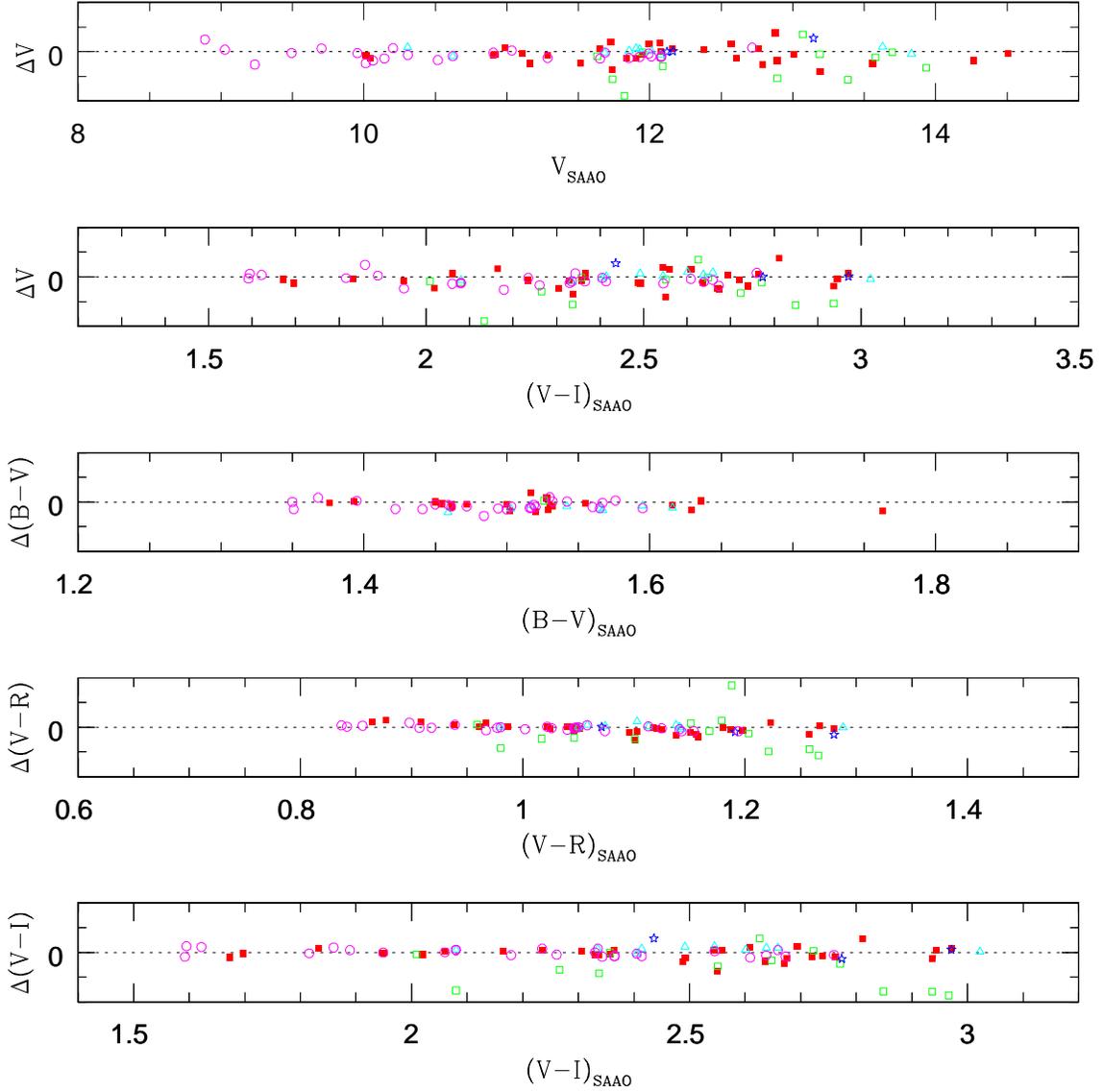}
\caption{ Comparison between our SAAO photometry and previously published data: solid squares
mark observations by Weis; open circles are from Bessell; open squares are from Eggen; triangles 
are from Leggett; and five-point stars are data from Patterson {\sl et al.}. The tick marks on
the vertical scale correspond to intervals of 0.1 magnitude. }
\end{figure}

\begin{figure}
\plotone{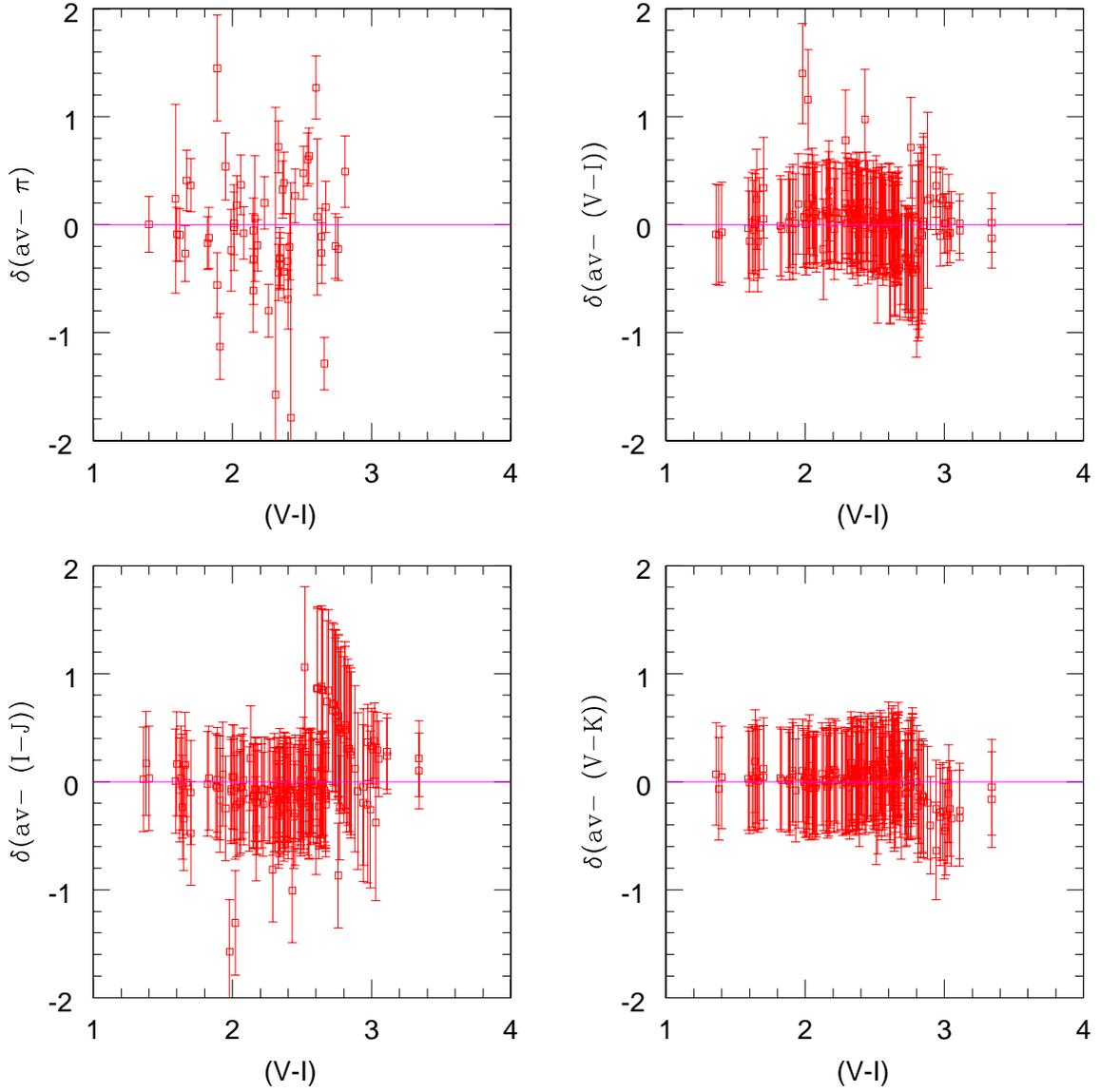}
\caption{ Residuals between the averaged distance modulus and individual estimators. plotted as 
a function of (V-I) colour. The increased residuals near (V-I)$\sim2.9$ reflect the steepening
in the main-sequence at that colour. }
\end{figure}

\begin{figure}
\plotone{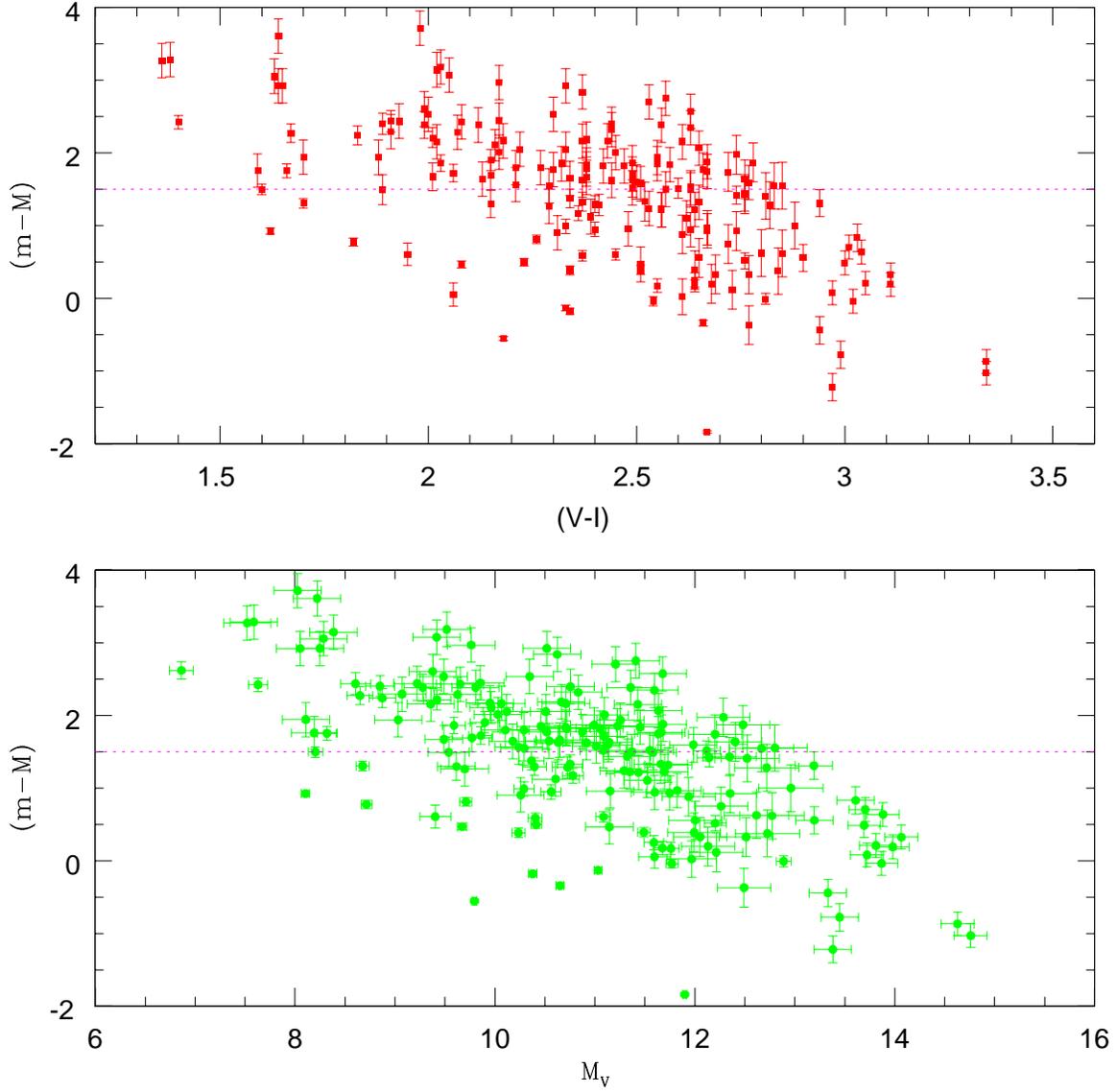}
\caption{ The distance distribution of the stars in Table 1 as a function of both (V-I) colour
and the inferred absolute magnitude. The dotted line marks a distance of 20 parsecs.}
\end{figure}

\end{document}